# Complex Eigenfrequencies of Some Axially Symmetrical Bodies

I. G. Efimova

Moscow State University

*Abstract.* The complex eigenfrequencies of a perfectly conducting cone, truncated cone, and finite circular cylinder are determined using the time-domain integral equation method and the Prony technique.

Nonstationary responses of scatterers to superwideband signals, in particular video pulses, are highly informative characteristics, which allows their application for identification of the geometrical structure of a body. When it is necessary to identify the shape of a body from a known set of configurations, the complex eigenfrequencies of a scatterer's exterior can serve as signatures [1].

As is known, at sufficiently late instants, when a free component dominates in a response, the transient field scattered by a perfectly conducting body can be represented as the series of complex exponents

$$E(t) = \sum_{i=1}^{\infty} A_i \exp(\gamma_i t),$$

where $\gamma_i$ are the complex eigenfrequencies of the body, which are the poles of the frequency-domain representation of the field, and $A_i$ are constants [2]. Numbers $\gamma_i$ are independent of the form of an incident signal, the spatial orientation of an object, or the observation angle and are determined by the object's shape [3].

Frequency- and time-domain methods can be applied to find eigenfrequencies $\gamma_i$. However, when the surface of a scatterer is such that the method of separation of variables cannot be employed, numerical frequency-domain techniques must be multiply applied to solve the direct problem of diffraction, which requires much CPU time. In this situation, the time-domain approach is more efficient, because the direct problem of nonstationary diffraction is solved only once, after which complex eigenfrequencies can be found from the obtained temporal response using the Prony method [4].

In this work, the effect of the shape of a body on its temporal response is analyzed for three types of bodies of revolution---a cone, a truncated cone, and a finite circular cylinder---illuminated by a Gaussian video pulse.
All of the bodies have the same maximum transverse dimension. The flare angles of the cone and truncated cone are equal to 23°. The ratio of the maximum transverse dimension to the longitudinal dimension is 0.505 for the cone, 0.749 for the truncated cone, and 0.5 for the cylinder.

The direct problem of nonstationary diffraction is solved using the time-domain integral equation method [5], which is one of the most universal and accurate numerical techniques. The temporal dependences of scattered fields are processed by the Prony method to obtain the complex eigenfrequencies of the aforementioned bodies. The calculations were performed for individual azimuthal harmonics (with the numbers $m = 0, 1, 2, 3$) of scattered signals and for the sum of several harmonics at various incidence angles of a primary signal and various observation angles.

As has been mentioned above, the complex eigenfrequencies of the exterior of an object are invariant to the spatial orientation of the object or the observation angle. However, the exact invariance is not actually observed. This can be attributed to the inaccurate approximation of a response by a finite sum of complex exponents and to the fact that the response contains the forced signal component, which has not decayed completely. The obtained values of complex eigenfrequencies were averaged over the sets of incidence and observation angles. The averaged values of $\gamma a/c$ (where $a$ is the maximum transverse dimension and $c$ is the velocity of light in free space) are summarized in the table.

Table. Complex eigenfrequencies ($\gamma a/c$) of various bodies of revolution

| $m$ | The shape of a scatterer | | |
| --- | --- | --- | --- |
| | Cone | Truncated cone | Cylinder |
| 0 | $-0.33 \pm i0.62$ | $-0.36 \pm i0.77$<br>$-0.78 \pm i1.62$ | $-0.27 \pm i0.55$ |
| | -0.62 ± i1.93<br>-0.67 ± i4.22<br>-0.53 ± i7.11 | -1.06 ± i2.67<br>-0.84 ± i3.97 | -1.03 ± i1.62<br>-1.05 ± i3.54<br>-0.92 ± i5.51 |
| 1 | -0.76 ± i0.65<br>-0.93 ± i1.04<br>-0.8 ± i2.09<br>-1.51 ± i3.36<br>-1.47 ± i6.75 | -0.81 ± i0.88<br>-0.66 ± i1.78<br>-0.16 ± i2.90<br>-0.62 ± i3.93 | -0.72 ± i1.00<br>-0.46 ± i1.93<br>-1.23 ± i3.97<br>-1.14 ± i6.69 |
| 2 | -0.75 ± i2.30<br>-0.81 ± i3.88 | -1.18 ± i2.48 | -0.91 ± i2.48 |
| 3 | -0.83 ± i3.91 | -1.21 ± i3.47 | -1.10 ± i3.28 |
| $\sum_{m=0}^{3}$ | -0.4 ± i0.58<br>-0.77 ± i2.61<br>-0.94 ± i4.17 | -0.29 ± i0.77<br>-0.65 ± i1.65<br>-1.46 ± i2.86 | -0.54 ± i0.58<br>-0.67 ± i2.04<br>-0.44 ± i3.34 |

The comparison of the averaged eigenfrequencies of a cone, truncated cone, and cylinder shows that each scatterer exhibits its own set of eigenfrequencies. The difference between these shapes is most distinctly revealed by the imaginary parts of the third and fourth (in the order of increasing Im($\gamma a/c$)) eigenfrequencies. Simulation has shown that the imaginary parts of complex eigenfrequencies are rather resistant to variations in the direction of the incident wave propagation and to variations in the observation angle. These imaginary parts can serve as signatures involved in identification of scatterers with close values of the ratio between the maximum longitudinal and transverse dimensions.